\documentclass[10pt,journal,compsoc]{IEEEtran}
\usepackage{ifpdf}

\usepackage{graphicx}
\usepackage[inline,shortlabels]{enumitem}
\usepackage{amsmath}
\usepackage{amssymb}
\usepackage{algorithm,algcompatible}
\usepackage[noend]{algpseudocode}
\usepackage{xcolor}

\ifCLASSOPTIONcompsoc
  \usepackage[nocompress]{cite}
\else
  \usepackage{cite}
\fi
\ifCLASSINFOpdf
\else
\fi

\usepackage{amsmath}

\ifCLASSOPTIONcompsoc
  \usepackage[caption=false,font=footnotesize,labelfont=sf,textfont=sf]{subfig}
\else
  \usepackage[caption=false,font=footnotesize]{subfig}
\fi

\usepackage{subfig}

\usepackage{url}

\begin{document}
%
\title{Availability-aware Service Placement Policy in Fog Computing Based on Graph Partitions}

%
%
%
%

\author{Isaac Lera, Carlos Guerrero, and Carlos Juiz,~\IEEEmembership{Senior Member,~IEEE}
        
\IEEEcompsocitemizethanks{\IEEEcompsocthanksitem  The authors are with the Computer Science Department, Balearic Islands University, Palma, SPAIN, E07122.\protect\\
Corresponding author: Carlos Guerrero
E-mail: carlos.guerrero@uib.es
}
}

%
%

\markboth{IEEE }%
{Shell \MakeLowercase{\textit{et al.}}: Bare Demo of IEEEtran.cls for Computer Society Journals}
%



\IEEEtitleabstractindextext{%
\begin{abstract}
This paper presents a policy for service placement of fog applications inspired on complex networks and graph theory. We propose a twofold partition process based on communities for the partition of the fog devices and based on transitive closures for the application services partition. The allocation of the services is performed sequentially by, firstly, mapping applications to device communities and, secondly, mapping service transitive closures to fog devices in the community. The underlying idea is to place as many inter-related services as possible in the most nearby devices to the users.  The optimization objectives are the availability of the applications and the Quality of Service (QoS) of the system, measured as the number of requests that are executed before the application deadlines. We compared our solution with an Integer Linear Programming approach, and the simulation results showed that our proposal obtains higher QoS and availability when fails in the nodes are considered. 
\end{abstract}

}

\maketitle

\IEEEdisplaynontitleabstractindextext

%
\IEEEpeerreviewmaketitle

\IEEEraisesectionheading{\section{Introduction}\label{sec:introduction}}

\IEEEPARstart{F}{og} computing has emerged as a suitable solution for the increase of application execution time and network usage that Internet of Things applications based on cloud services generate. This paradigm establishes that the in-network devices are provided with computational and storage capacities, and it enables them to allocate or execute services of the IoT applications that are commonly executed in the cloud provider~\cite{openfog2017openfog}. By this, the application services are placed closer to the users (or IoT) devices and, consequently, the network latency between users and services and the network usage are reduced. Nevertheless, the limited capacities of the in-network devices, also known as fog devices in this domain, make the definition of management policies even more necessary than in other distributed system such as cloud computing.

The objective of our work is to study an application service placement policy to maximize service availability in case of failures. The placement consist on the selection of the most suitable fog devices to map service instances. We consider that the IoT applications are defined as a set of interrelated services that are initially and permanently deployed on the cloud provider, but that they can be horizontally scaled by creating new stateless instances in the fog devices. We also consider that the users of our domain are unalterable connected to a same gateway or access point, i.e., we consider that our users are IoT devices such as sensors or actuators, instead of considering mobility patterns, as for example in the case of mobile users.

We propose a two phases policy that is addressed to optimize the service availability, in terms of reachability of the services from the IoT devices, and the deadline satisfaction ratios, in terms of the percentage of requests that obtain the application responses before their deadlines. In the first phase, the policy maps applications (the complete set of interrelated services) to a set of well-connected devices to guarantee the availability of the application for the users connected to that set. We propose to use the community structure of the fog devices for the generation of the partitions of those devices. Once that an application is mapped to a fog community, a second allocation process is performed, by mapping the services of the application to the fog devices in the community. This second phase addresses the optimization of the response time by prioritizing the allocation of interrelated services in the same fog device. We propose to partition the services of an application by using the transitive closure of a service to determine the services to be placed together in the same device.

Fog service placement problem has been addressed in previous researches, even considering community-based approaches~\cite{8377095}, but we address some features that have not been previously considered, and the novel contributions of our approach are:

\begin{itemize}
	\item The combination of the use of complex network communities for the device partition and service transitive closures for the application partition, that has not been used simultaneously in previous studies.
	\item The optimization of both the application deadline satisfaction, considered in some previous studies, and the application availability, not included in previous studies, and their evolution along the simulation.
	\item An experimental validation that includes dynamic fails of the infrastructure along the simulation.  
\end{itemize}

\section{Related Work}
\label{relatedwork}

The problem of the optimization of service placement in a fog architecture has been previously addressed from several different prespectives, by considering algorithm proposals such as genetic algorithms~\cite{7867735,Skarlat2017}, Montecarlo methods~\cite{7919155}, distributed solutions~\cite{Guerrero2018lightweight}, Petri Nets~\cite{7935527}, Markov processes~\cite{URGAONKAR2015205}, and being linear programming one of the most common solutions~\cite{7359164,Velasquez2017,HUANG201447,7110527, 7511465, 7422054}.

Nevertheless, there is still room for improvement and some research challenges have not been still covered. For example, most of the previous solutions have included the optimization of response time, power consumption, cost, or network usage. But to the best of our knowledge, they have not studied the availability and the influence of failures in the infrastructure.

The use of the community relationship of the devices of a distributed system for the optimization of the resource management was initially proposed by Filiposka et al.~\cite{Filiposka2015}, and they applied it in the optimization of the allocation of virtual machines in a datacenter to optimize the hop distances between related virtual machines. In the field of fog computing, the use of other topological features of graphs and complex network was proposed at a later stage, such as centrality indexes for the static definition of fog colonies~\cite{guerrero2018influence} or the placement of data in fog devices~\cite{Lera2018comparing}.

The idea of organizing the complex  structure of a fog architecture have been applied in several studies, where the authors defined these static infrastructure organizations as fog colonies~\cite{Skarlat2017}, micro-clouds~\cite{7867723}, Foglets~\cite{Bonomi2014}, or fog domains~\cite{8246720}. For example, Skarlat et al.~\cite{Skarlat2017} defined a twofold distributed placement policy that first considered if a service should be allocated in a fog colony or migrated to the neighbor colony. Once that the colony was chosen, the control node of the colony decided the device that allocated the service. In all those studies, the partition of the fog devices was static and unique for all the applications.

On the contrary, Filiposka, Mishev and Gilly proposed a virtual partition of the devices that is specific for each application and it is dynamically established by the conditions of the system. They implemented an evolution of the proposal in~\cite{Filiposka2015} for the case of allocation of virtual machines (VM) into fog devices~\cite{8377095}. They considered that the fog services where encapsulated in one VM and they proposed a two phases optimization process, where in the first step the VM is mapped to a device community, and in the second step, the VM is allocated in any of the devices in the community with a traditional optimization technique. This is probably the most similar work to our proposal in terms of the optimization algorithm, but with a different optimization objective. Their objective was to propose a run-time algorithm for the migration of the VM as mobile user of the applications move through different access points to reduce the average service delay.

The main differences of the work of Filiposka et al. with our proposal are: first, we study the suitability of the community relationships to improve service availability instead of the migration of VMs due to the user mobility; second, we consider a more complex structure of the applications because we defined them as a set of interrelated services that can be allocated in different devices, while they defined the applications as a single encapsulating element, the VM; third, we also study the use of a graph partitioning approach, the transitive closure of the services, for the allocation of the services inside the communities to also benefit the placement of the most interrelated services in the same devices to reduce the network delays between interrelated services.

\section{Problem Statement}
\label{sec:problem}

\begin{figure}
	\centering
	\includegraphics[width=0.35\textwidth]{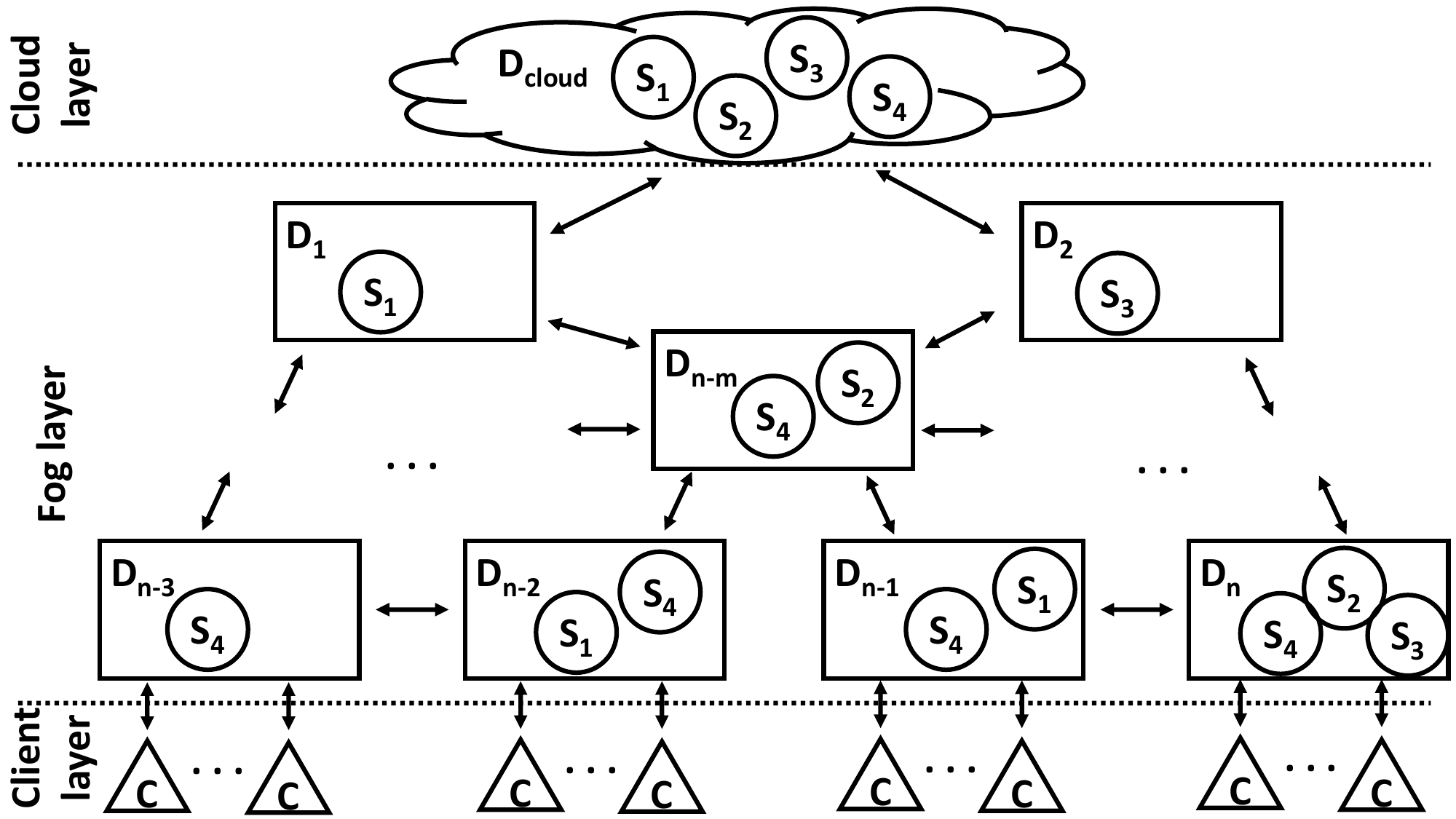}
	\caption{Fog computing architecture.}
	\label{fig:fogarchitecture}
\end{figure}

A general fog computing architecture is represented in Fig.~\ref{fig:fogarchitecture} where three layers can be identified: cloud layer, fog layer and client layer. Three types of devices can be differentiated: a device for the cloud provider of the cloud layer; the gateways, that are the access points for the clients; the fog devices, the network devices between the cloud provider and the gateways. All the devices have resources to allocate and execute services. 

The fog infrastructure can be modeled as a graph where the nodes are the devices and the edges the direct network links between devices.  We identify those devices as $D_i$, considering two special cases for the cloud provider ($D_i^{cloud}$) and the gateways  ($D_i^{gtw}$). The devices are defined by the available capacity of their resources $AR_{D_i}$, that is a vector which contains the capacities of each physical component. For the sake of simplicity, we have considered a scalar value, but it could easily be extended
by including as many elements as necessary. We suppose unlimited resources for the specific case of the cloud provider, $AR_{D_i^{cloud}}=\infty$. The devices are also defined by the processing speed $IPT_{D_i}$, measured in term of instructions per unit of time. The network links are identified by the two connected nodes $NL_{D_i,D_j}$, and we consider that it is a bidirectional communication, $NL_{D_i,D_j} = NL_{D_j,D_i}$. The network links are defined by the propagation delay, $PR_{NL_{D_i,D_j}}$, and the network bandwidth, $BW_{NL_{D_i,D_j}}$. Thus, the network delay, $ND_{NL_{D_i,D_j}}$, for the transmission of a packet between two connected devices is calculated as:
\begin{equation}
\label{eq:networkdelay}
ND_{NL_{D_i,D_j}} = PR_{NL_{D_i,D_j}} + \frac{size}{BW_{NL_{D_i,D_j}}}
\end{equation} 
where $size$ is the size of the packet to be transmitted.

The applications in our problem domain follow a microservice based development pattern, that is increasingly being used in IoT applications~\cite{Vogler:2016:SFP:2909066.2850416,7300793,Saurez2016}. This type of applications are modeled as a set of small and stateless services that interoperate between them to accomplish a complex task~\cite{7436659}. Thus, the services can be easily scale up, by downloading the encapsulating element and executing it, or scale down, by just stopping and removing instances of the service. We assume that there is at least one instance of each service running in the cloud provider ($D_i^{cloud}$). 

We model each application $APP_x$ as a directed graph, where the nodes are the services and the edges are the request messages between the services. We identify the services as $S_u$ and they are defined by the resource consumption generated in the device that allocates the service, $CR_{S_u}$. As in the case of the available resources in a device, the resource consumption is generally defined as a vector which measures the consumption of each physical component, but we have considered a scalar value for a simpler definition of the problem. Services are executed when a request message is received. We classify the services in two types depending on the origin of the service request: the entry-point service $S^{ep}_u$, the origins of the request messages that arrive to those services are users $US_a$ or IoT devices (sensors typically) $ID_b$; the intra-services $S^{intra}_u$, that are only requested by other services. An intra-service can be requested for several different services and the entry-point service can be requested for several users or IoT devices. But, we suppose that there is only one entry-point service for each application.

The task performed by a service is different depending on the requester, so the execution generated by a request not only depends on the service but also on the requester, i.e. the request message. The request messages are identified by the origin and target services,  $MS_{S_u,S_v}$, and they are modeled as unidirectional edges, $MS_{S_u,S_v} \neq MS_{S_v,S_u}$. The requests generated by the users or the IoT services, i.e. the requests to the entry-point services, are only identified by the target entry-point service $MS_{\emptyset,S_u}$.

The request messages are defined by the size of the request message $SZ_{MS_{S_u,S_v}}$, that determines the transmission time of the service request, and the execution load that the target service will generate in the device, defined by the number of instructions to be executed, $EI_{MS_{S_u,S_v}}$.

We assume that there is at least one instance of each service in the cloud provider. But those services can be horizontally scaled by deploying new instances in the fog devices. By this, the workload can be distributed between instances and the network delay from the user to te service is reduced. We define a placement matrix, $P$, of size $|S_u| \times |D_i|$, number of services per number of fog devices, where a element $p_{ui}$ is equal 1 if service $S_u$ is deployed in device $D_i$, and 0 otherwise.

The placement of the services are constrained by the device resource capacity. The resources consumed by the allocated services should not exceed the available resources in the device:
\begin{equation}
\label{eq:resourceusage}
\sum_{u=1}^{|S_u|} \left( p_{ui} \times CR_{S_u} \right) \le AR_{D_i},\ \forall\ D_i
\end{equation}

Our optimization objectives are to increase the application deadline satisfaction ratio, and the application availability as the devices or the network links fail.

We define the deadline satisfaction ratio as the percentage of application requests that are processed before the application deadline. Consequently, the applications in the system, $APP_x$, need to be defined by their deadlines, $DL_{APP_x}$. The user perceived response time, $RT_{RQ^n_{US_a,APP_x}}$, is the metric that measures the time between a specific application request is sent by the user ($RQ^n_{US_a,APP_x}$) and all the application services finish their execution. It includes the network delay of the request between services and the response times (execution and waiting time) of the services.

The equation for the deadline satisfaction ratio is:
\begin{equation}
\label{eq:deadline}
deadline(US_a,APP_x) = \frac{|RT_{RQ^n_{US_a,APP_x}} < DL_{APP_x}|}{|RQ^n_{US_a,APP_x}|}
\end{equation}
where $|RQ^n_{US_a,APP_x}|$ is the number of times that a a request for $APP_x$ is sent from user $US_a$, and $|RT_{RQ^n_{US_a,APP_x}} < DL_{APP_x}|$ is the number of those requests that satisfied the application deadline. This metric can be generalized by considering the request to an application from any user, $deadline(APP_x)$, or the ratio for all the applications and users in the system, $deadline(system)$.

Our second objective, the application availability, is defined as the ratio of users that are able to reach all the services of the applications they request for a given point in time. In a hypothetic case, where any of the elements in the system fails, the service availability would be 1.0. But the devices or the network links can fall down, breaking the shortest paths between the users and the application services. At best, this only would generate an increase in the network delay due to the requests would be routed by a longer path, damaging the deadline satisfaction ratios. But it could even result in making the user impossible to reach all the application services, damaging the service availability ratio. The equation for the service availability ratios is:
\begin{equation}
\label{eq:availability}
availability(APP_x) = \frac{|US_a,\ g.t.\ \exists\ path\ US_a\ to\ APP_x|}{|US_a,\ g.t.\ US_a\ requests\ APP_x|}
\end{equation}
 
In summary, our domain problem is addressed to find $P,\ p_{ui}\ \forall\ S_u,D_i$ by minimizing $deadline(US_a,APP_x) \wedge (1 - availability(APP_x))\ \forall\ US_a,APP_x$ subject to the constraint in Eq.\eqref{eq:resourceusage}.

\section{Two Phases Partition-based Optimization Proposal}

Our optimization algorithm is based on a two phases placement process with a first mapping of applications in fog communities and a second phase which allocates the services of an application in the devices of a fog community. We partition the fog devices using the community relationship of the complex network that models the network infrastructure of the system.  The application services are partitioned considering the transitive closures of the nodes that represent the services in the application graph. 

We study if the community relationships of the fog devices is a good indicator to detect device sets that guarantee the availability of the services and the reachability of the devices when device and network links failures are considered. Additionally, we also study if the transitive closure of a service is a good indicator to decide the services that are allocated in the same device to avoid network communications overheads. 

\subsection{Community-based Fog Devices Partition}

The first phase of our optimization algorithm deals with the mapping between applications (a set of interrelated services) and a device partitioning. We propose to partition the devices with the use of the community relatioship between them. This phase of our optimization algorithm is based on the previous work of Filiposka, Mishev and Gilly, where they studied and validated community-based algorithms for placement optimization in cloud computing~\cite{Filiposka2015} and in fog computing~\cite{8377095}.

The community structure is a topological feature of graphs that determines the sets of nodes which are better connected between them than with the rest of the network. The most popular community detection method is the one proposed by Girvan and Newman~\cite{PhysRevE.69.026113}, which detects communities by progressively removing edges from the original graph. The algorithm removes the edges with the highest betweenness centrality, at each step. Betweenness centrality of an edge is the sum of the fraction of the shortest paths that pass through the edge. Therefore, a community, that is organized with two regions that are mainly communicated by only one edge, is split into two new communities in each algorithm iteration. 

Under the conditions of our domain problem, a device community can be understood as a set of devices that are well connected between them, with alternatives communication paths, and that the shortest paths between devices are evenly distributed between the topology. Consequently, a fail in an edge inside the community will have a lower influence in the communication paths between devices than a fail in the edges that connect the communities. This lower influence means that the fails inside the communities will not generate isolated regions in the topology neither an important increase in the communication delays.

The Girvan-Newman method iteratively determines the communities and the dendrogram, the tree structure of the communities, can be built. We characterized those communities with its  depth in the dendrogram. We define this depth as the iteration in which the community was obtained. The higher the depth value is, the better communicated the device community is. Consequently, from the point of view of the availability, it is better to place the applications in device communities with higher depth values, since the devices inside those communities are better communicated between them than the devices in communities with lower depths values~\cite{PhysRevE.70.056104}.

\begin{figure}
	\centering
	\includegraphics[width=0.35\textwidth]{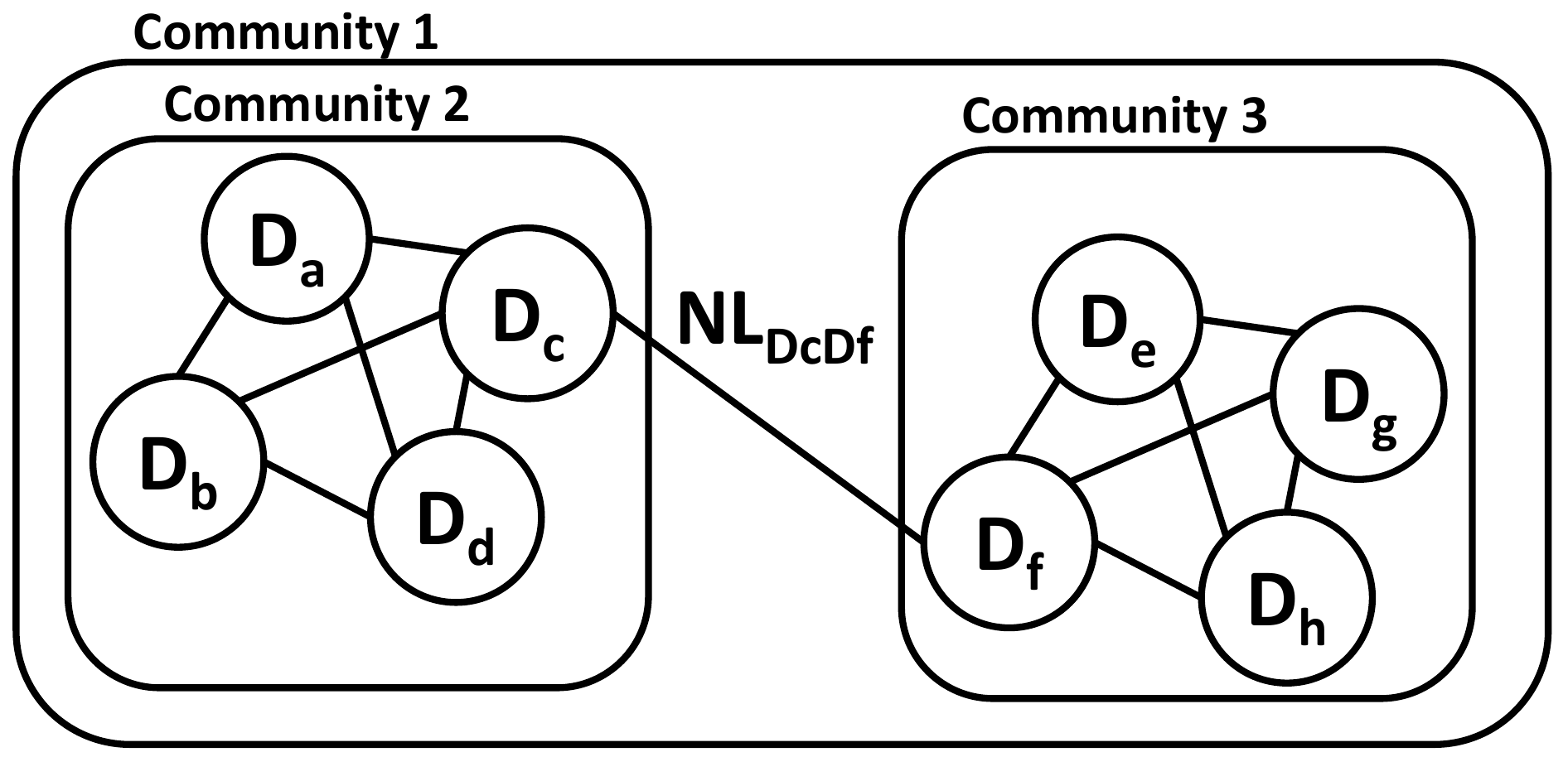}
	\caption{Example of fog device communities.}
	\label{fig:community}
\end{figure}

For example, consider the fog infrastructure in Fig.~\ref{fig:community}. The network link $NL_{D_c,D_f}$ is the one with the highest edge betweenness centrality since it is passed through the highest number of shortest paths. If we iterate the Girvan-Newman method over this example, communities 2 and 3 have higher depth values than community 1 since they are obtained when $NL_{D_c,D_f}$ is removed in the next iteration of the community generation algorithm. Consider also that we deploy an application with services $S_i$ and $S_j$ in community 1, allocating $S_i$ in $D_a$ and $S_j$ in $D_h$, and that the user that requests the application is connected to device $D_b$. Under those conditions, a fail in $NL_{D_c,D_f}$ would  make impossible to finish the execution of the application since their services are unreachable. On the contrary, if we deploy the application in community 2, any fail in a edge would not make impossible to execute the application. Finally consider that a second user is connected to device $D_h$. The best alternative, from the point of view of the availability, would be to horizontally scale up by deploying the same application twice in both communities 2 and 3, than only once in any of them.

This example shows that, in an unrealistic situation with unlimited resources in all the devices, the best option would be to deploy an instance of the application for each client that requests it and this deployment would be placed in the community with the highest depth value that includes the device where the client is connected to. But this cannot be performed due to the limited resources in the devices of a community. Moreover, if we note that the higher the depth value of the community, the smaller the number of devices in the community, i.e., the communities with the highest values are the ones formed by only one device. Consequently, it is necessary to prioritize the allocation of the applications in the communities. We propose to use a greedy algorithm for this priorization, more concretely, the First-Fit Decreasing algorithm~\cite{6822306}. 

Our optimization algorithm deals, in this first step, with the placement of applications in device communities using a First-Fit Decreasing approach. The priority criteria for ordering the applications is their execution deadlines, by prioritizing the applications with shortest deadlines. The algorithm starts checking the allocation of the application from the device communities with highest depth to the ones with the lowest, and the application is allocated in the first community with enough resources to allocate all the services of the application. If after checking all the communities, the application has not been allocated, this will be available only in the cloud provider. The process for the same application is repeated as many times as the number of users in the system that request this application. Algorithm~\ref{appallocation} shows the pseudo-code of our proposal. The algorithm goes through the applications (in ascending deadline order), the users that request them and the communities (in descending depth order), trying to allocate the services of the application in the devices in the community. 

In this first step, we map the applications in communities, but the map of services remains to be defined. We separate the process in two steps because we mainly focus the first one  (mapping applications to communities) on increasing the application availability, and the second one (mapping services of an application to devices in a device community) on the application deadlines. This second step is performed by the function \textit{placeServicesInDevices()}, in line~\ref{line:placeServicesInDevices}, and its details are explained in Section~\ref{applicationcommunities} and Algorithm~\ref{servallocation}.

Our algorithm checks if an application has been previously placed in a community (line~\ref{line:alreadyplaced}), and if not, it delegates the decision to place the application to the community to the algorithm which ckecks if the application services fit into the device community (Algorithm~\ref{servallocation}).

\begin{algorithm}[t]
	\caption{Device community-based application allocation}
	\label{appallocation}
	\scriptsize
	\begin{algorithmic}[1]
		\State $\mathbb{C} \leftarrow$ calculate device communities
		\State $\mathbb{IC} \leftarrow$ order communities $\mathbb{C}$ by descending depth
		\State $\mathbb{A} \leftarrow$ order applications by ascending deadline
		\State \textit{appPlacement} $\leftarrow\ \emptyset$ 
		\For{ \textit{app} \textbf{in} $\mathbb{A}$}
		\State $\mathbb{U} \leftarrow$ get users requesting application \textit{app}
			\For{ \textit{user} \textbf{in} $\mathbb{U}$}
				\State \textit{dev} $\leftarrow$ get device where \textit{user} is connected
				\For{ \textit{infCom} \textbf{in} $\mathbb{IC}$}
					\If {\textit{dev} $\in$ \textit{infCom}} 
						\If {\textit{infCom} $\in$ \textit{appPlacement}[\textit{app}]} \label{line:alreadyplaced}
							\State \textit{""application app already placed in community infCom""}
							\State \textbf{break}
						\Else
							\If {placeServicesInDevices(\textit{app},\textit{infCom})} \label{line:placeServicesInDevices}
								\State \textit{appPlacement}[\textit{app}].append(\textit{infCom})
								\State update resource usages in \textit{infCom}
								\State \textit{""placed application app in community infCom""}
								\State \textbf{break}
							\EndIf
						\EndIf					
					\EndIf
				\EndFor
			\EndFor
		
		\EndFor

	\end{algorithmic}
\end{algorithm}

\subsection{Transitive Closure-based Application Partition}
\label{applicationcommunities}

Once that the mapping of a given application into a candidate community of devices is performed by the first phase of the optimization algorithm, the second phase deals with the allocation of the services of the application into the devices in the community. We first partition the applications into sets of services, and it is checked if each of those service sets can be placed in just one device. If not, smaller sets are considered. The partition of the service into sets is based on our previous work~\cite{Guerrero2018lightweight}, where we studied and validated the use of a distributed placement algorithm where the service sets are created by considering the transitive closure of the services in the application graph.

The transitive closure of a directed graph indicates the nodes that are reachable for each of the nodes in the graph. If a vertex $j$ is reachable by a vertex $i$ means that there is a path from $i$ to $j$. The reachability matrix of a graph is called the transitive closure of the gragh, and the set of reachable nodes for a given node is called the transitive closure of a node~\cite{warren1975modification}.

Under the conditions of our domain problem, the transitive closure of a node can be understood as the set of services that are requested for the execution of the given service, i.e., the outgoing requests generated by a service when it receives an incoming request. If we are interested in reducing the response time of the application execution, the services of the transitive closure should be allocated in the same device to reduce the communication delays between them, since the network delay is 0.0 for request messages inside the same device. Moreover, the best case is when all the services of an application are allocated in the same device, but this is limited by the resource constraint  (Equation~\ref{eq:resourceusage}). 

We also propose a First-Fit algorithm for this second phase (Algorithm~\ref{servallocation}), which orders the sets of services from the ones with the biggest sizes (only one transitive closure with all the services) to the smallest sets of services (the transitive closures with only one node or with the loops in the service flow), and tries to place those sets of services into a same device. The devices are ordered by a fitness value which is the theoretical user perceived response time. This value is obtained by adding the network latency between the device and the user and the execution time of all the services in the device. This prioritize the devices that are both closer to the users and faster in the execution. By this, the second step of the algorithm optimizes the user perceived response time, and, consequently, improves the deadline satisfaction ratio.

\begin{figure}
	\centering
	\includegraphics[width=0.45\textwidth]{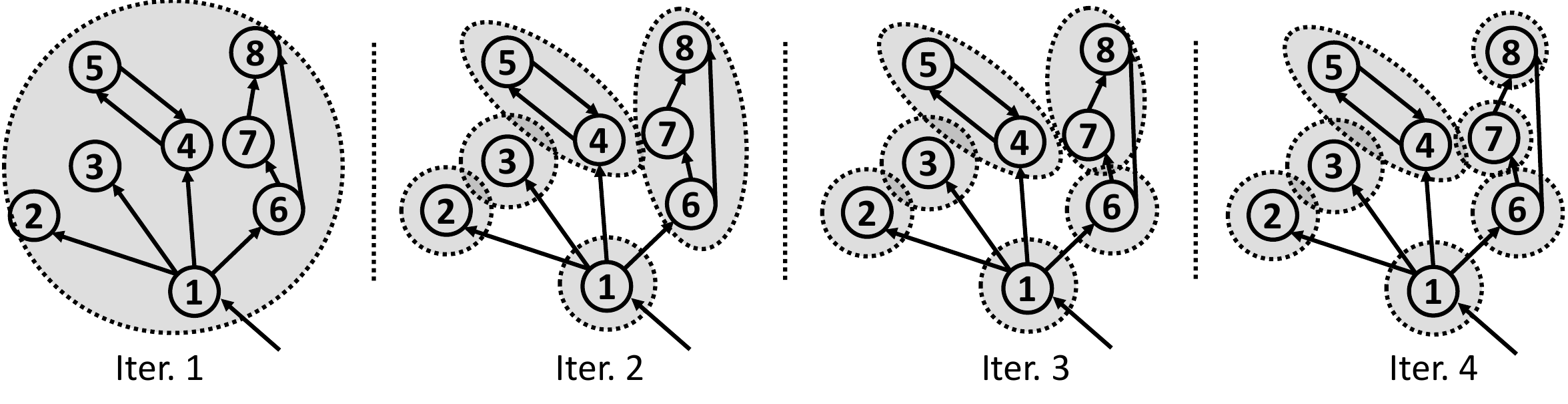}
	\caption{Example of service transitive closures.}
	\label{fig:transitiveclosure}
\end{figure} 


Initially, Algorithm~\ref{servallocation} goes through the devices ordered by the fitness value, and tries to allocate as much services as possible in the devices with the highest values. For the first device, it first tries to allocate all the services of the application. If they do not fit, the service set is split in several sets, one for each entry-point service and one additional set for the transitive closures of each of its neighbor services of the entry-point one, and it checks if any of those new sets fits in the first device. This is recursively repeated for each transitive closure set that contains services not previously allocated. Fig.~\ref{fig:transitiveclosure} shows an example of how the transitive closure of the services is generated along the iterations of the algorithm that partition the services of the application.  

Once that all the service sets have been evaluated to be placed in the first device, this process is sequentially repeated for all the devices for the unallocated services. If after considering all the devices, there are still unallocated services, the mapping of the application in the current device community is rejected. Consequently, the first phase of the algorithm has to consider a greater community for the placement.

\begin{algorithm}[t]
	\caption{Transitive closure-based service allocation}
	\label{servallocation}
	\scriptsize
	\begin{algorithmic}[1]
				\Function{placeServicesInDevices}{}
		\State $\mathbb{TC} \leftarrow$ generate transitive closure partitions for \textit{app}
		\State $\mathbb{D} \leftarrow$ order devices in \textit{infCom} by reponse time
		\State $\mathbb{SP} \leftarrow \emptyset$ \COMMENT{\ \ \ \  /* Services already placed*/}
		\State \textit{servPlacement} $\leftarrow\ \emptyset$ 
		\For{ \textit{dev} \textbf{in} $\mathbb{D}$}
		\For{ \textit{appPartition} \textbf{in} $\mathbb{TC}$}
			\For{\textit{closure} \textbf{in} \textit{appPartition} }
			    \If {(closure \textbf{not in} $\mathbb{SP}$) \textbf{and} (\textit{closure} fits \textbf{in} \textit{dev})}
				
					\State $\mathbb{SP} = \mathbb{SP}\ \cup$ \textit{closure}
					\For {\textit{service} \textbf{in} \textit{closure}}
					\State \textit{servPlacement}[\textit{service}] = \textit{dev}
					\EndFor
					\State update resource usages in \textit{dev}
					\If {$\mathbb{SP}$ == \textit{app}}
					\State \Return \textbf{True}, \textit{servPlacement}
					\EndIf

			    \EndIf

			\EndFor

				\EndFor
			
		\EndFor
		\State \Return \textbf{False}, $\emptyset$
		\EndFunction
	\end{algorithmic}
\end{algorithm}

\section{Experimental Evaluation}



\begin{table}[!t]
	\renewcommand{\arraystretch}{1.3}
	\caption{Values of the parameters for the experiment characterization}
	\label{table_expparameters}
	\centering
	\begin{tabular}{|c|l||r|}
		\hline
		\textbf{Parameter} & & \textbf{min.--max.}\\
		\hline
		\hline
		
		\multicolumn{1}{|l|}{\textbf{Network}} &&\\
		\hline
		 Propagation time (ms) & $PR_{NL_{D_i,D_j}}$& 5\\
		\hline
		Bandwidth (bytes/ms)  & $BW_{NL_{D_i,D_j}}$& 75000\\
		\hline
		\multicolumn{1}{|l|}{\textbf{Fog device}} &&\\
		\hline
		 Resources (res. units) & $AR_{D_i}$& 10--25\\
		\hline
		Speed (Intrs/ms) & $IPT_{D_i}$& 100--1000\\
		\hline
		\multicolumn{1}{|l|}{\textbf{Application}} &&\\
		\hline
		 Deadline (ms)&$DL_{APP_x}$& 300--50000\\
		\hline
		Services (number)&&2--10 \\
		\hline
		Resources (res. units)&$CR_{S_u}$& 1--6\\
		\hline
		Execution (Intrs/req)&$EI_{MS_{S_u,S_v}}$& 20000--60000\\
		\hline
		Message size (bytes)&$SZ_{MS_{S_u,S_v}}$& 1500000--4500000\\
		\hline
		\multicolumn{1}{|l|}{\textbf{IoT device}} &&\\
		\hline
		 Request rate (1/ms)&&1/1000--1/200 \\
		\hline
		Popularity (prob.)&& 0.25\\
		\hline
	\end{tabular}
\end{table}

%

We defined random characteristics for the elements of our simulation experiments. We modeled the parameters of the elements in the domain with uniform distributions and the minimum and maximum values are shown in Table~\ref{table_expparameters}.

The service applications were generated randomly following a growing network (GN) graph structure. GN graphs are built by adding nodes one at a time with a link to one previously added node. The network infrastructure was created as a random Barabasi-Albert network with 100 fog devices.
Betweenness centrality index is a topological metric that measures the number of shortest path that goes through a device. 
The gateway devices were selected from the nodes placed in the edges of the network, i.e., the nodes with the smallest betweenness centrality indices. Betweenness centrality index is a topological metric that measures the number of shortest path that goes through a device. We selected the 25\%  of devices with the lowest centrality value to behave as gateways (25 gateways). The number and the applications requested from the IoT devices connected to the gateways were determined with a popularity distribution modeled with an uniform distribution.

The random experimental scenario finally resulted on 20 applications with 106 services, that totally needed 360 resource units and the fog devices were able to offer up to 1874 resources units. 70 IoT devices (or users) were deployed and they generated an application request each 1/557 ms in average.

We compared the results of our proposal with the ones obtained from the implementation of an integer linear programming (ILP) service allocation optimizer. As we mention in Section~\ref{relatedwork}, ILP solutions are the most numerous in fog service placement optimization. 

The experiments were executed using the YAFS simulator that we had previously developed for other research works. This simulator is able to include graph-based network topologies and pluggable fog service placement policies, apart from other features that, to the best of our knowledge, are not provided by other fog simulators, such as node failures, or dynamic service placement and routing. The simulator is open source and it can be downloaded from its code repository~\cite{yasf}.

The experiment results are presented and analyzed in two separated sections. Section~\ref{simulresults} includes the analysis of the results obtained with the YAFS simulator. Those results compare the user perceived response time and the availability of the applications for the IoT devices. In Section~\ref{placementresults}, it is presented an analysis of the service placement obtained with both optimization policies (our proposal and the ILP one).

\begin{figure*}[!t]
	\centering
	\includegraphics[width=0.80\textwidth]{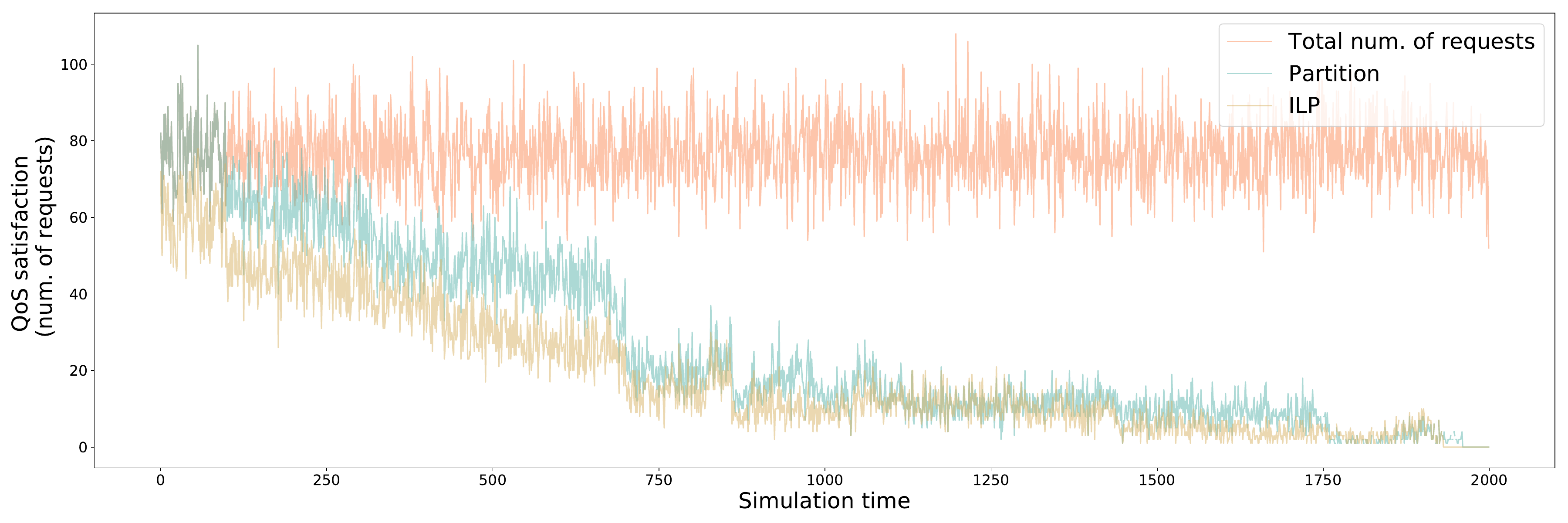}
	\caption{Evolution of the QoS with regard to the fail of fog devices, in terms of the number of requests which satisfy application deadlines ($|RT_{RQ^n_{US_a,APP_x}} < DL_{APP_x}|$) compared with the total number of requests ($|RQ^n_{US_a,APP_x}|$).}
	\label{fig:QoSevolution}
\end{figure*} 

\begin{figure}
	\centering
	\includegraphics[width=0.35\textwidth]{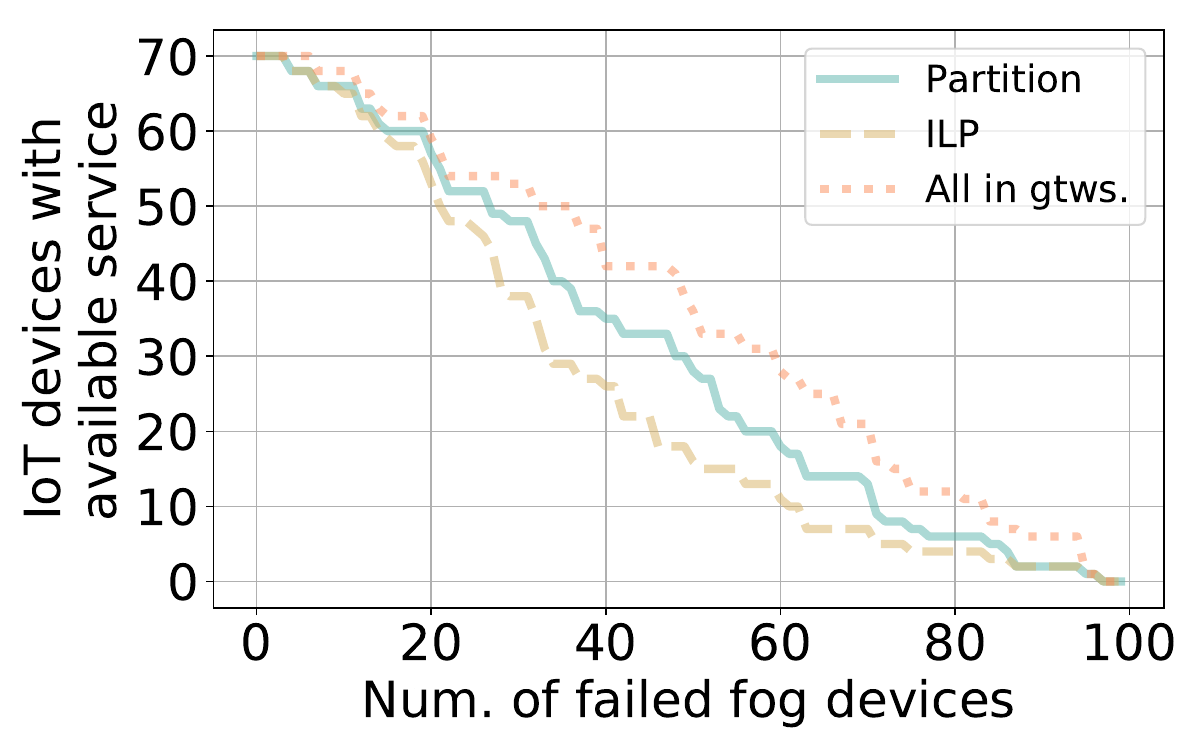}
	\caption{Number of IoT devices that get services in regard with the number of failed fog devices ($availability(APP_x)$).}
	\label{fig:availableapps}
\end{figure} 

\subsection{Simulation Results}
\label{simulresults}

A first simulation scenario included fails in the fog devices to study the availability of the services when the nodes are getting down. The simulation included random and permanent fails in the nodes, starting with all the devices (100 nodes) alive, and finishing the simulation with fails in all of them. The fails were generated uniformly along the simulation. The results of this simulation are presented in Fig.~\ref{fig:QoSevolution} and shows the QoS in terms of the total number of requests that are executed satisfying the application deadline. 
The reason because a request does not satisfied the deadline can be both due to the response time is higher than the deadline or due to none device with the services of the requested application are reachable from the IoT device due to all the paths between them have failed devices. Three data series are represented in Fig.~\ref{fig:QoSevolution}: one for the total number of requests that are sent from the IoT devices (labeled with \textit{Total num. of requests}), one for the number of requests that are executed before the deadline when the placement of our solution is considered (labeled with \textit{Partition}); and the number of requests that satisfied the deadline with the ILP policy (labeled with \textit{ILP}).

It is observed that our approach results in a higher number of satisfied requests, mainly during the first half of the simulation (up to 50 failed devices). In the second part of the simulation, improvements in the QoS are also observed but these are less significant in regard with the ILP. 

For the sake of a deeper analysis of the availability, it has been also measured in terms of the number of IoT devices that are able to request their applications thank to that all the services they need are reachable with network paths without failed devices. This is represented in Fig.~\ref{fig:availableapps}, where the y-axis are the number of IoT devices that are able to request their applications, and the x-axis the number of devices that have failed. The figure also includes the hypothetic and impossible case, due to the resource limit constraint, of allocating all the services in the gateways (labeled as \textit{All in gtws.}). This is the best case and is useful to compare the solutions with the best upper bound.  These results confirm that our proposal is able to increase the availability of the system when fails happen in the fog devices.

\begin{figure*}
	\centering
	\includegraphics[width=0.9\textwidth]{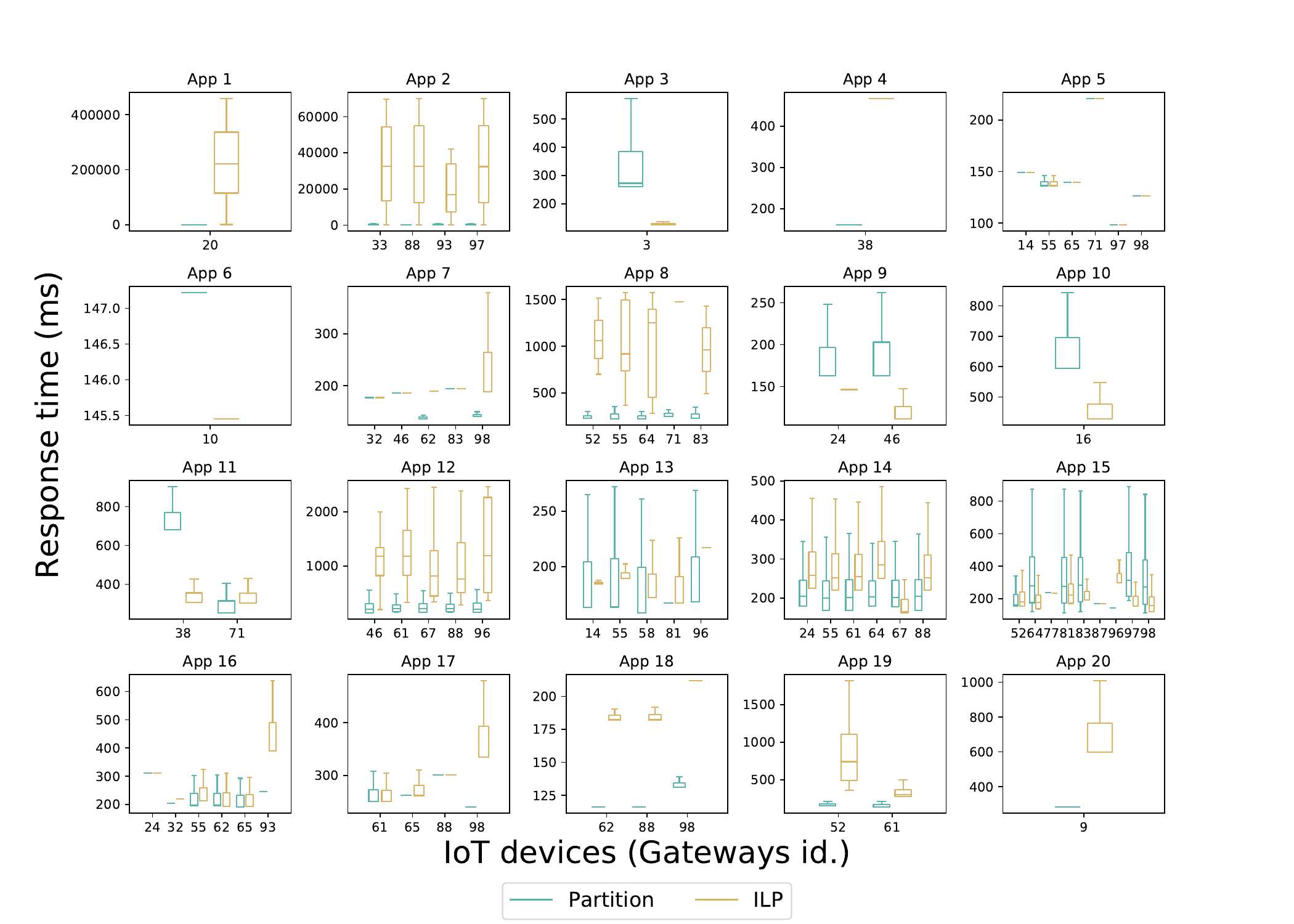}
	\caption{User perceived response times of the applications for each user (or IoT device) in the system ($RT_{RQ^n_{US_a,APP_x}}$).}
	\label{fig:executiontime}
\end{figure*}

A second simulation scenario did not included fails in the fog devices and was used to study the user perceived response time of the applications. These response times were measured as the time between the user request was generated in the IoT device and all the application services finished. The results were measured independently for each pair application-IoT device.  They are summarized in Fig.~\ref{fig:executiontime}. Each plot in the figure represents the response times of an application, an each item in the x-axis corresponds to one gateway that has an IoT device (or user) that request the application. The results of our solution are labeled as \textit{Partition} and the results of the ILP approach are labeled as \textit{ILP}.

It is observed that the placement obtained with our proposal does not reduce the response time for all the applications, but it is shorter for 13 of the 20 applications. Additionally, we can observed that in some applications an important damage of the response time is obtained. This is explained because both policies prioritize applications with shorter deadlines in front of the ones with longer deadlines. Nevertheless, there are less of these extreme cases, and with shorter times, when our policy is used: our policy only damages application 15 with a time of around 1000 ms, in front of four applications up to 400 s with the ILP policy (around 400000 ms for application 1, 300000 ms for application 8, 200000 ms for application 12, and 70000 for application 2).

In summary, our service placement policy shows a better behavior in terms of availability of the services that also results on a better QoS in the system. On the contrary, the response time of some applications results damaged but this behavior is also observed with the ILP policy, generating even worse response times.

\begin{figure*}[!t]
	\centering
	\subfloat[Allocation of the services in the fog devices ($P,\ p_{ui}\ \forall\ S_u,D_i$).]{\includegraphics[width=0.3\textwidth]{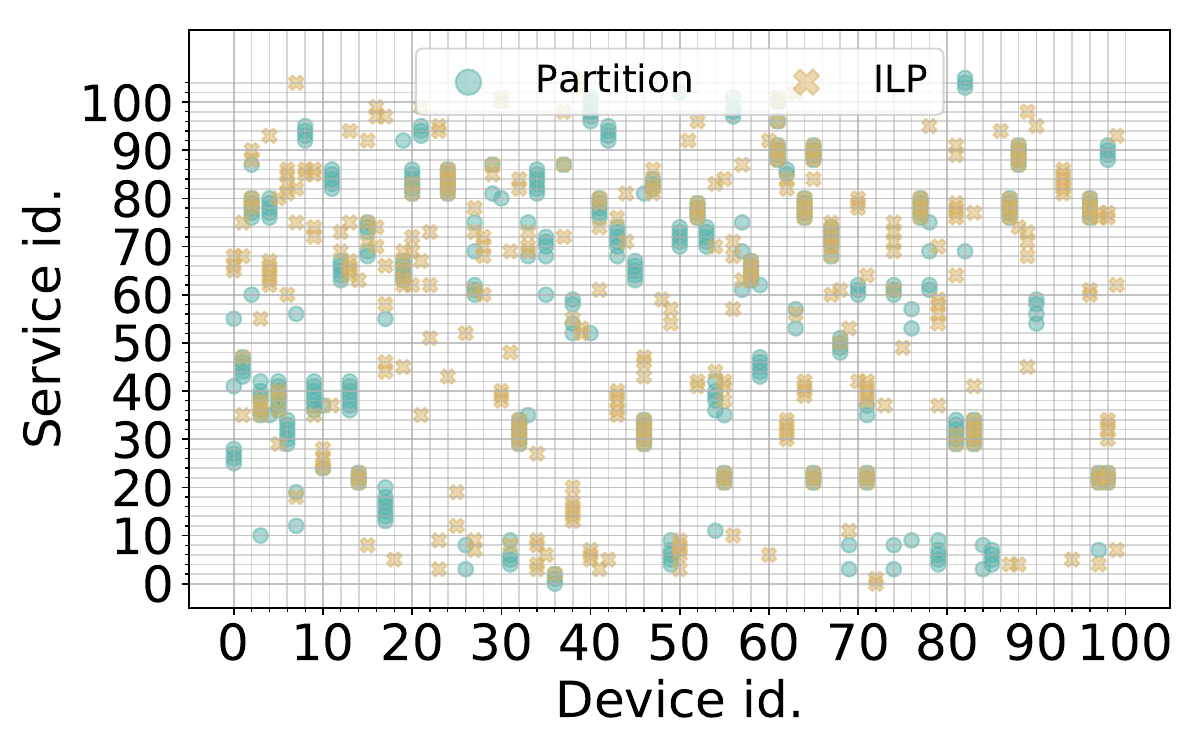}%
		\label{fig:allocation}}
	\hfil
	\subfloat[Resource usage of the fog devices ($\sum_{u=1}^{|S_u|} \left( p_{ui} \times CR_{S_u} \right), \forall D_i$).]{\includegraphics[width=0.3\textwidth]{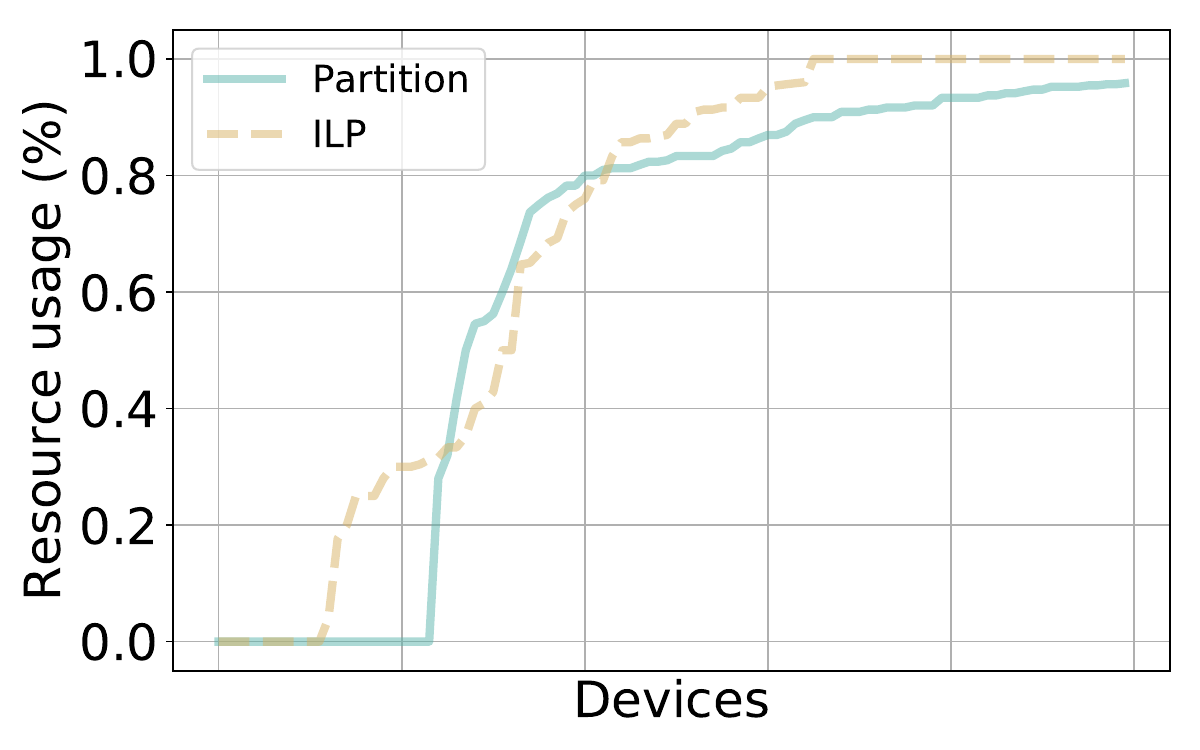}%
		\label{fig:resourceusage}}
	\hfil
	\subfloat[Service allocation in terms of hop distance with the IoT devices.]{\includegraphics[width=0.3\textwidth]{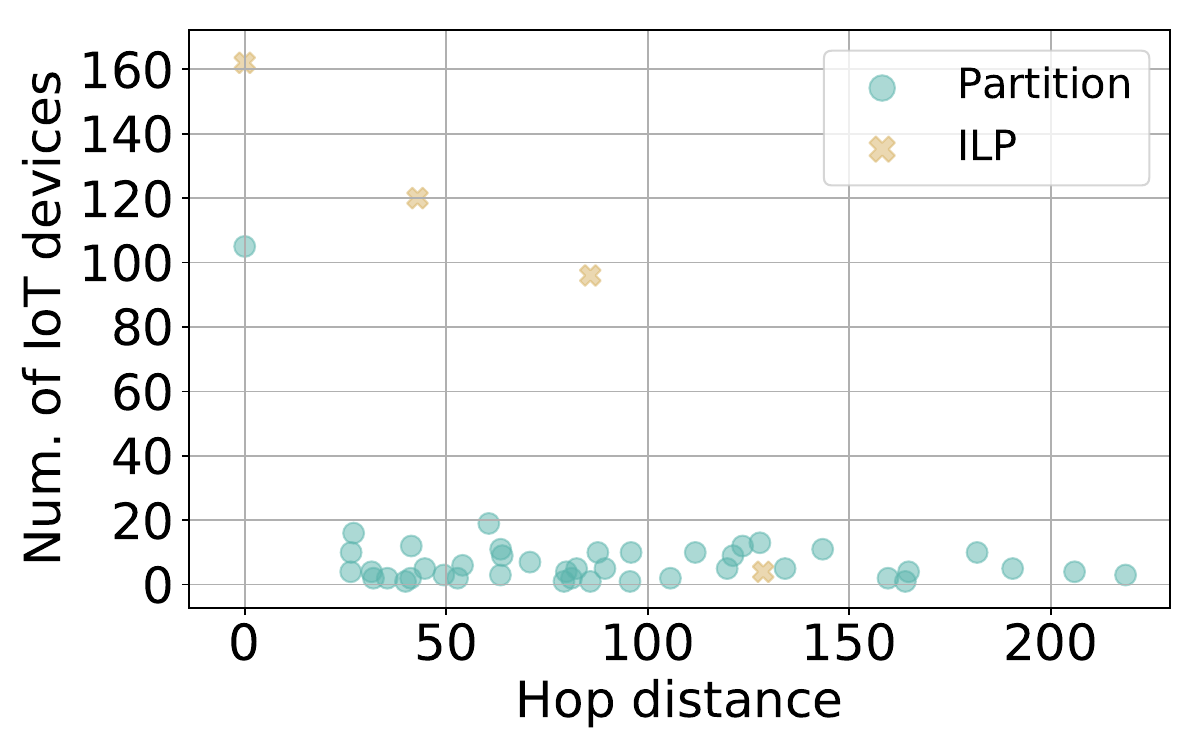}%
		\label{fig:hopdistance}}
	
	\caption{Comparison of the services placement between our partition-based algorithm and the ILP optimizer.}
	\label{placementcomparison}
\end{figure*}

\subsection{Placement Results}
\label{placementresults}

This section is devoted to compare the placement of the services obtained from the execution of our algorithm with regard to the ILP one. This analysis is included to give a brief idea of how the services are spread across the fog devices. 

Firstly, Fig.~\ref{fig:allocation} shows that the placement of the services differs a lot between both placement policies. A mark in the plot of the figure indicates that a given service (y-axes) is placed in a given device (x-axes). Taking into account that the services of the same application have consecutive identifiers, it is also observed that in the case of our policy (\textit{Partition}), there are more cases of devices that allocate several services of the same application (consecutive marks in the same device).

Fig.~\ref{fig:resourceusage} represent the resource usage of the fog devices. The y-axis represents the percentage of resources that are used by the services allocated in a given device and the x-axis are the devices ordered by these percentages in ascending order. By the analysis of the figure, we can observe that in the placement of our solution, there are almost the double of nodes that do not allocate any service (the resource usage is 0.0), and there is not any device that is fully used (resource usage of 1.0), with regard to the case of the ILP where  almost 40 devices have a 100\% usage of the resources. The first interpretation of these results is that the scale level of our solution is smaller than the ILP one, in fact, we calculated that our policy deployed 357 (and 1161 resource units) instances of the services and the ILP deployed 374 (and 1203 resource units), around 5\% more services (3.6\% more resources). Consequently, our solution is able to obtain better QoS and availability with a lower use of the fog resources (smaller number of instances). The second interpretation is that the services are more evenly distributed, since the workload of the devices is smaller, avoiding the saturation of the devices and keeping the system in a more flexible state in order to allocate new service instances.

Finally, Fig.~\ref{fig:hopdistance} shows the relationship between the service placement and the hop distance between the allocated service and the IoT device that requests it. A point in the scatter plot indicates how many IoT devices has a given distance with a service of the application they request. For example, in the case of our policy, there are around 100 services that are allocated in the gateways where the IoT devices are connected (a hop distance of 0.0). On the contrary, the ILP policy allocates more than 160 services in the gateways, the point (0,160) in the plot. We observe that the services are distributed more evenly and placed further from the gateways (higher distances) for the case our policy. Consequently, the ILP is able to place the services closer to the IoT devices. Despite this, our policy shows a better general behavior also in terms of application response time.

%
%
%
%

%

\section{Conclusion}

We have proposed an algorithm for service placement in fog devices based on the partition of the fog devices (into communities) and the services of the applications (into transitive closures) for the optimization of the QoS of the system and the service availability for the users (or IoT devices).

Two simulation scenarios have been executed, one including fails in the fog devices and another one without fails, to measure the response time of the applications, the service availability and the number of request that were served satisfying the application deadlines. The service placement obtained with our policy resulted in a higher QoS and service availability, with regard to the placement of an ILP-based algorithm. In the case of the user perceived response time, our policy obtained better times for 13 of the total 20 applications. Both policies showed a high degradation of service for some applications, but in the case of the ILP, 
this degradation happened in more applications and resulting in longer response times.

As future works, the use of complex networks and graph theory for the optimization of other parameters of the systems, such as service cost, network usage, migration cost, and service provider cost could be studied. By the own nature of the proposed policy, the optimization of these other metrics probably would need to be combined with other type of heuristics to obtain suitable results, and consequently, further research is necessary.

\ifCLASSOPTIONcompsoc
\section*{Acknowledgments}
\else
\section*{Acknowledgment}
\fi

This research was supported by the Spanish Government (Agencia Estatal de Investigaci\'on) and the European Commission (Fondo Europeo de Desarrollo Regional) through grant number TIN2017-88547-P (MINECO/AEI/FEDER, UE). 



%


\ifCLASSOPTIONcaptionsoff
  \newpage
\fi



\bibliographystyle{IEEEtran}
\bibliography{mybibfile}
%



%

\vspace{-10 mm}

\begin{IEEEbiography}[{\includegraphics[width=1in,height=1.25in,clip,keepaspectratio]{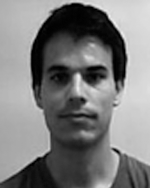}}]{Isaac Lera}
	received his Ph.D. degree in Computer Engineering at the Balearic Islands University in 2012. He is an assistant professor of Computer Architecture and Technology at the Computer Science Department of the University of the Balearic Islands. His research lines are semantic web, open data, system performance, educational innovation and human mobility. He has authored in several journals and international conferences.\end{IEEEbiography}

\vspace{-15 mm}

\begin{IEEEbiography}[{\includegraphics[width=1in,height=1.25in,clip,keepaspectratio]{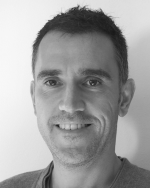}}]{Carlos Guerrero}
received his Ph.D. degree in Computer Engineering at the Balearic Islands University in 2012. He is an assistant professor of Computer Architecture and Technology at the Computer Science Department of the University of the Balearic Islands. His research interests include web performance, resource management, web engineering, and cloud computing. He has authored around 40 papers in international conferences and journals.\end{IEEEbiography}

\vspace{-15 mm}

\begin{IEEEbiography}[{\includegraphics[width=1in,height=1.25in,clip,keepaspectratio]{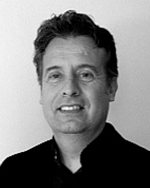}}]{Carlos Juiz}
received his Ph.D. degree in Computer Engineering at the Balearic Islands University in 2001. He is an associate professor of Computer Architecture and Technology at the Computer Science Department of the University of the Balearic Islands. His research interests include performance engineering, cloud computing and IT governance. He has authored around 150 papers in different international conferences and journals.\end{IEEEbiography}




\end{document}